\documentclass[aps,prd,preprint,nofootinbib,a4paper,11pt,superscriptaddress]{revtex4-1}

\usepackage[T1]{fontenc}
\usepackage[centertags]{amsmath}
\usepackage{amsfonts}
\usepackage{amssymb}
\usepackage[sort&compress]{natbib}
\usepackage[hypertex]{hyperref}
\usepackage[dvips]{graphicx}

\DeclareMathOperator{\Sp}{Sp}
\DeclareMathOperator{\re}{Re}
\DeclareMathOperator{\im}{Im}

\newcommand{\lan}{\langle}
\newcommand{\ran}{\rangle}
\newcommand{\spp}{\mathbf{p}}
\newcommand{\spx}{\mathbf{x}}

\newcommand{\e}{\varepsilon}
\newcommand{\vf}{\varphi}
\newcommand{\s}{\sigma}

\newcommand{\al}{\alpha}
\newcommand{\be}{\beta}
\newcommand{\ga}{\gamma}
\newcommand{\Ga}{\Gamma}
\newcommand{\de}{\delta}

\newcommand{\la}{\lambda}
\newcommand{\La}{\Lambda}

\begin{document}


\title{Gravitational mass-shift effect in the standard model}

\date{\today}

\author{P.O. Kazinski}
\email[E-mail:]{kpo@phys.tsu.ru}
\affiliation{Physics Faculty, Tomsk State University, Tomsk, 634050 Russia}

\begin{abstract}

The gravitational mass-shift effect is investigated in the framework of the standard model with the energy cutoff regularization both for stationary and non-stationary backgrounds at the one-loop level. The problem concerning singularity of the effective potential for the Higgs field on the horizon of a black hole, which was reported earlier, is resolved. The equations characterizing the properties of the vacuum state are derived and solved in a certain approximation for the Schwarzschild black hole. The gravitational mass-shift effect is completely described in this case. The masses of the massive particles in the standard model are shown to depend on the value of the Higgs boson mass in the flat spacetime. If the Higgs boson mass in the flat spacetime is less than 263.6 GeV, then the mass of any massive particle approaching a gravitating object grows. If the Higgs boson mass in the flat spacetime is greater than or equal to 278.2 GeV, the masses of all the massive particles decrease in a strong gravitational field. The Higgs boson masses lying in between these two values prove to lead to instability, at least at the one-loop level, and so they are excluded. It turns out that the vacuum possesses the same properties as an ultrarelativistic fluid in a certain approximation. The expression for the pressure, the entropy and enthalpy densities of this fluid are obtained. The sound speed in this fluid is also derived.

\end{abstract}


\maketitle

\section{Introduction}

Recently, many attempts have been undertaken to verify the prediction of general relativity in its classical formulation. One of these important predictions is the gravitational redshift law. Despite that it was directly tested with a high accuracy \cite{Vessot}, several programs are now planned to check the redshift law in the weak field limit to find possible smaller deviation \cite{ACES}. Small deviations from the standard redshift law arise in the models that do not preserve the local Lorentz-invariance. The latter is believed to be violated on the Planck scale. In this paper we are about to investigate a somewhat different way of possible violations of the standard redshift law. It results from the gravitational mass-shift effect by means of the Higgs mechanism \cite{olephfsb}.

The standard derivation of the redshift law in the course of general relativity heavily relies on the assumption that the emission spectrum does not depend on the external gravitational field in the reference frame associated with the emitter. However, if the masses of particles (say, the electron mass) change with gravitational field, the spectrum will also change even in this system of coordinates. The masses of all massive particles of the standard model are generated via the Higgs mechanism and determined by a nonzero vacuum expectation value of the Higgs field. This expectation value provides the minimum to the effective potential, while the shape of this potential turns out to depend on the background gravitational field \cite{olephfsb}. This leads to the gravitational mass-shift effect and, as a consequence, to deviations from the standard redshift law. This is a purely quantum effect and it stems from the dependence of the zero-point energy of quantum fields on the background gravitational field.

One of the crucial steps in evaluation of the zero-point energy is a correct definition of the Hamiltonian of the whole system in the Hilbert space. The formal expression for the Hamiltonian diverges and needs an appropriate regularization procedure. Another equivalent formulation of this problem is that we should define the normal ordering for the composite operators \cite{Lowenst,BogolShir}. For a stationary gravitational field, we have a distinguished set of the creation-annihilation operators which are associated with the stationary mode functions. The latter are defined as eigenfunctions of the Lie derivative along the Killing vector related to stationarity (for details, see, e.g., \cite{DeWpaper}). This allows us to define the normal ordering and prescribe a rigorous meaning to composite operators such as the Hamiltonian. As soon as the well-defined Hamiltonian is given, the vacuum state is also defined as the state with minimum energy (in black hole physics this vacuum is called the Boulware vacuum \cite{Boulware}). This approach is equivalent to the so-called physical regularization \cite{Collins}, which, loosely speaking, consists in modification of the particles dispersion laws above a certain cutoff energy in such a way that the loop diagrams become convergent. Both regularization procedures require a Killing vector. It is the square of this Killing vector that enters the effective potential of the Higgs field, changing the shape of the potential in the presence of external gravitational field. For the flat spacetime, this Killing vector ``disappears'' from the effective action (as its square is a mere constant) and, in the one-loop approximation, the effective potential takes a familiar form of the Coleman-Weinberg potential \cite{ColWein}. So, after renormalization, there is no Lorentz-invariance violation for the flat spacetime according to this approach.

In the non-stationary case, we have to introduce some vector field $\xi^\mu$ that allows us to define the physical regularization (the notion of a dispersion law) and the normal ordering just as in the stationary case. This vector field must coincide with the Killing vector in the case of a stationary background. It turns out that this vector field is uniquely defined by the requirement of covariant divergenceless of the matter energy-momentum tensor, provided that the system starts its evolution from some stationary state. The equations describing the evolution of $\xi^\mu$ have a hydrodynamic form. In particular, there exists a conserved charge, which can be interpreted as the entropy of the whole system. As we shall see, in a reasonable approximation, these equations can be cast into the form of the Euler equations for an ultrarelativistic fluid with the equations of state determined by the effective potential of the Higgs field. This will allow us to find stationary solutions to these equations for the Schwarzschild black hole. There are two such solutions. The first one is the trivial solution coinciding with the Killing vector. It corresponds to the case of a stable star in a certain approximation, of course. The second solution is nontrivial and describes an ``accretion'' of the vacuum onto the black hole. In this case the effective potential turns out to be non-singular on the horizon of the black hole. The masses of massive particles of the standard model acquire a finite shift and do not vanish there. In general, the gravitational mass-shift effect is greater for stable stars than for black holes of the same mass at the same distance from the gravitating object.

As long as the vector field $\xi^\mu$ enters the effective potential of the Higgs field, it spoils the so-called local position invariance (see, e.g., \cite{Will}). Hence, the approach we are going to study can be somewhat considered among the Lorentz-invariance violating models. Let us emphasize the distinctions with the standard methods of introducing the Lorentz-invariance violation \cite{Bjork,Coleman,Kostel,Odints,AACGS}. The vector field discussed above must always be a background field, i.e., it is not a quantum field or its average. It characterizes the regularization procedure and necessarily appears in the course of defining the composite operators in the Hilbert space. Whatever new particles are introduced into the model, this vector field has to be included to define the physical regularization and accomplish the theory. The equations of motion of $\xi^\mu$ do not follow from the action principle and arise as a self-consistency condition of the model. This approach is minimal in the following sense: As we have already mentioned, the Lorentz-invariance is not violated for the flat spacetime after renormalization; There is no additional degrees of freedom or structures in the model. In the stationary case, the vector field $\xi^\mu$ is a time-like Killing vector and determined by the metric. In the non-stationary case, this vector field is uniquely defined by the self-consistency condition. This minimality will allow us to describe the gravitational mass-shift effect and make certain predictions rather than just to fit various parameters of the model to the experimental data.

The paper is organized as follows. In Sec. \ref{StandMod}, we introduce a notation and recall the basic features of the standard model that are necessary for deriving the effective action. As we restrict ourself to the one-loop approximation, we need only the spectrum of masses of the real and fictitious particles of the standard model. It is obtained in this section for the Feynman gauge.

In Sec. \ref{EffPot}, we derive the one-loop effective potential of the Higgs field on a stationary background in a certain approximation. It has the same form as the effective potential of the Higgs field on the Schwarzschild background for the $O(N)$-$\phi^4$ model \cite{olephfsb}. The different regularization prescriptions and the dependence of the final result on them are also discussed in this section. The form of the effective potential obtained is completely fixed by imposing three normalization conditions. Further, the gravitational mass-shift effect on a stationary background is considered for different values of the Higgs boson mass. As expected, there are two scenarios of the mass behavior at small the Killing vector squared. The first one corresponds to the infinitely growing masses of massive particles at small $\xi^2$. The second scenario describes the decreasing masses. In the latter case, the symmetry of the standard model is restored at sufficiently small the Killing vector squared. Realization of these cases depends on the sign of the undetermined constant resulting from the regularization procedure. The sign of this constant depends, in its turn, on the value of the Higgs boson mass in the flat spacetime. This sign changes at the ``critical'' value of the Higgs boson mass which is approximately equal to $278.2$ GeV. If the Higgs boson mass is less than this critical value, the first scenario is realized. In the opposite case and, in particular, at the critical value of the Higgs boson mass, the second scenario takes place.

Section \ref{NonStatCase} is devoted to the case of a non-stationary background. It begins with derivation of the general equations describing an evolution of the vector field $\xi^\mu$ discussed above. Then, in a certain approximation, these equations are reduced to the Euler equations for an ultrarelativistic fluid. The temperature of this fluid is proportional to $(\xi^2)^{-1/2}$, i.e., to the Tolman temperature \cite{Tolman} on a stationary background. Therefore gravity effectively heats such fluid. Further, the speed of sound for this fluid is introduced. It characterizes the velocity of propagation of small disturbances of the vector field $\xi^\mu$. In the weak field limit, this sound speed is independent of particular details of the standard model and its square equals the Newtonian potential with reversed sign (the first cosmic velocity). The limit of the sound speed at small $\xi^2$ is also universal and equals $1/\sqrt3$, as it should be for a fluid at the very high temperature, although this situation seems not to be realized. In a general case, the entropy of this fluid, or the pressure, or the speed of sound cannot be found analytically for an arbitrary $\xi^2$ in terms of elementary functions. However, if the Higgs boson mass is equal (or close) to the critical value, this can be done with a high accuracy for those $\xi^2$ that are realized in practice. The formula for the sound speed proves to be independent of the details of the standard model in this case too. In the interval of the Higgs boson masses from $263.3$ GeV to the critical value $278.2$ GeV, the square of the speed of sound becomes negative for certain $\xi^2$ and the system is hydrodynamically unstable there. We exclude this region of the Higgs boson masses from our subsequent considerations. Section \ref{NonStatCase} is concluded by consideration of the stationary solutions to the equations of motion of the vector field $\xi^\mu$ on the Schwarzschild background. Noteworthily, the problem of a spherically symmetric accretion admits of an analytic solution for the Higgs boson masses close to the critical value.

We use the metric tensor $g_{\mu\nu}$ with the signature $-2$ and the system of units in which $\hbar=c=1$. The Greek indices are raised and lowered by this metric. The gravitational field is assumed to be fixed and non-fluctuating. An inclusion of its quantum fluctuations at the one-loop level does not influence the results.

\section{One-loop correction to the effective action}\label{StandMod}

Let us briefly recall some basic features of the standard model that will be relevant for our consideration (see for details \cite{Carringt,Kapusta,Okun}). The generally covariant action of the standard model can be divided into several parts
\begin{equation}
    S_{\text{SM}}=\int d^4 x\sqrt{|g|}\mathcal{L}_{\text{SM}},\qquad\mathcal{L}_{\text{SM}}=\mathcal{L}_{\text{gauge fields}}+\mathcal{L}_{\text{leptons}}+\mathcal{L}_{\text{quarks}}+\mathcal{L}_{\text{Yukawa}}+\mathcal{L}_{\text{Higgs}},
\end{equation}
where the Lagrangian densities for the gauge fields and leptons are
\begin{equation}
\begin{gathered}
  \mathcal{L}_{\text{gauge fields}}=-\frac14 f_{\mu\nu}f^{\mu\nu}-\frac14 F_{\mu\nu}^aF^{\mu\nu}_a-\frac14F_{\mu\nu}^\al F^{\mu\nu}_\al,\\
  f_{\mu\nu}=\partial_{[\mu}B_{\nu]},\qquad F_{\mu\nu}^a=\partial_{[\mu}A^a_{\nu]}+g\e^a_{bc}A^b_\mu A^c_\nu,\qquad F_{\mu\nu}^\al=\partial_{[\mu}A^\al_{\nu]}+g_sf^\al_{\be\ga}A^\be_\mu A^\ga_\nu,\\
  B\in u(1),\qquad A^a\tau_a\in su(2),\qquad A^\al\la_\al\in su(3),
\end{gathered}
\end{equation}
and
\begin{equation}
  \mathcal{L}_{\text{leptons}}=\sum_{i=1}^3\biggl[\bar{l}^i_L(i\hat{\partial}+\frac{g}{2}\hat{A}^a\tau_a-\frac{g'}2\hat{B})l^i_L+\bar{l}^i_R(i\hat{\partial}-g'\hat{B})l^i_R\biggr].
\end{equation}
Here $l^i_L$ are the left-handed $SU(2)$ doublets, $l^i_R$ are the right-handed $SU(2)$ singlets, and $g_s$, $g$ and $g'$ are the coupling constants. The covariant derivative acting on the spinors is defined in the standard way
\begin{equation}
    \hat{\partial}=\gamma^\mu(\partial_\mu+\frac18\omega_{\mu ab}[\ga^a,\ga^b]),\qquad\ga^\mu=\ga^ae^\mu_a,
\end{equation}
with $e^\mu_a$ and $\omega_{\mu ab}$ being the tetrad and spin connection, respectively. The quark sector reads
\begin{equation}
\begin{split}
  \mathcal{L}_{\text{quarks}}&=\sum_{i=1}^3\biggl(\overline{\left[
                                \begin{array}{c}
                                  u_i \\
                                  d'_i \\
                                \end{array}
                              \right]}_L(i\hat{\partial}+\frac{g_s}2\la^\al \hat{A}_\al+\frac{g}{2}\hat{A}^a\tau_a+\frac{g'}6\hat{B})
                              \left[
                                \begin{array}{c}
                                  u_i \\
                                  d'_i \\
                                \end{array}
                              \right]_L+\\
                              &+\bar{u}^i_R(i\hat{\partial}+\frac{g_s}2\la^\al \hat{A}_\al+\frac{2g'}3\hat{B})u^i_R+
                              \bar{d}'^i_R(i\hat{\partial}+\frac{g_s}2\la^\al \hat{A}_\al-\frac{g'}3\hat{B})d'^i_R\biggr),\\
                              d'^i&=U_{\text{CKM}}^{ij}d^j,\qquad u^i=(u,c,t),\quad d^i=(d,s,b),
\end{split}
\end{equation}
where $U_{\text{CKM}}$ is the Cabibbo-Kobayashi-Maskawa matrix. The masses of the real massive particles of the standard model are generated by the terms
\begin{equation}
  \mathcal{L}_{\text{Yukawa}}=-\frac1{\sqrt{2}}\sum_{i=1}^3\biggl(f_i^u\overline{\left[
                                \begin{array}{c}
                                  u^i \\
                                  d^i \\
                                \end{array}
                              \right]}_L\phi u^i_R+f_i^d\overline{\left[
                                \begin{array}{c}
                                  u^i \\
                                  d^i \\
                                \end{array}
                              \right]}_L\phi_c d^i_R+f^l_i\bar{l}^i_L\phi l^i_R+\text{h.c.}\biggr),
\end{equation}
where $f_i$ are the Yukawa couplings and
\begin{equation}
  \mathcal{L}_{\text{Higgs}}=\frac12|(i\partial_\mu+\frac{g}{2}A^a_\mu\tau_a+\frac{g'}2 B_\mu)\phi|^2-\frac{\mu^2}2|\phi|^2-\frac{\la}4|\phi|^4,
\end{equation}
where $\phi$ is the $SU(2)$ doublet, $\mu^2$ is a negative constant, and $\la$ is the Higgs self-interaction coupling constant.

We consider the minimal standard model, that is, the neutrinos entering into the left-handed $SU(2)$ doublets $l^i_L$ are assumed to be massless and the possible non-minimal coupling $\bar{\xi} R|\phi|^2$ of the Higgs field with gravity is set to zero. As we shall see, the particles with small masses do not considerably change the effective potential and, consequently, do not affect the gravitational mass-shift effect. As for the non-minimal coupling term, it can be combined with the term $\mu^2|\phi|^2$ on the Einstein spaces \cite{Petrov}. So, as long as we neglect the back-reaction of the matter on the metric, this term can be omitted. It should be also noted that if we allow for the gravitational field $g_{\mu\nu}$ to fluctuate and integrate out the gravitons, the effective Higgs mass is changed by a quantity that depends on $\bar{\xi}$ (see, for example, \cite{BezShap,Starob}). However, on the electroweak scale, this mass-shift is considerable only for the enormous values of the non-minimal coupling $\bar{\xi}$, where the very applicability of the perturbation theory becomes questionable. Hence, we disregard this term.

We shall study the effective action of the Higgs field under the assumption that the vacuum expectation values of other fields are zero. To this end, we parameterize the Higgs doublet as
\begin{equation}
    \phi=\left[
           \begin{array}{c}
             0 \\
             \eta+\chi \\
           \end{array}
         \right]+i\zeta^a\tau_a
         \left[
           \begin{array}{c}
             0 \\
             1 \\
           \end{array}
         \right]=\left[
           \begin{array}{c}
             \zeta_2+i\zeta_1 \\
             \eta+\chi-i\zeta_3 \\
           \end{array}
         \right],
\end{equation}
where $\zeta_a$ are the Goldstone bosons, $\eta$ is a vacuum expectation value of the Higgs field, and $\chi$ describes the fluctuations of the Higgs field. Then the in-out effective action takes the form
\begin{equation}
\begin{split}
    e^{i\Ga[g_{\mu\nu},\eta]}&=\int\limits_{\text{1PI}} D\Phi DcD\bar{P} e^{iS_{\text{tot}}[g_{\mu\nu},\eta,\Phi,c,\bar{P}]},\qquad\left[
                                                                                                                                    \begin{array}{c}
                                                                                                                                      0 \\
                                                                                                                                      \eta \\
                                                                                                                                    \end{array}
                                                                                                                                  \right]=\frac{\lan\text{out}|\phi|\text{in}\ran}{\lan\text{out}|\text{in}\ran},
    \\ S_{\text{tot}}&=S_{\text{SM}}+S_{\text{gauge fixing}}+S_{\text{ghosts}},
\end{split}
\end{equation}
where $\Phi$ is the full set of fields including the Goldstone bosons and $\chi$, while $c$ and $\bar{P}$ are the Faddeev-Popov ghosts. The notation for the path-integral above means that only the one-particle irreducible diagrams are taken into account. The states $|\text{in}\ran$ and $|\text{out}\ran$ denote in- and out-vacua. They differ only by a phase for stationary backgrounds. In the Feynman gauge we shall use, the gauge fixing and ghosts actions are given by
\begin{equation}
\begin{gathered}
  \mathcal{L}_{\text{gauge fixing}}=-\frac12(\nabla^\mu A_\mu^a-\frac12 g\eta\zeta^a)^2-\frac12(\nabla^\mu B_\mu+\frac12g'\eta\zeta^3)^2-\frac12(\nabla^\mu A_\mu^\al)^2,\\
  \mathcal{L}_{\text{ghosts}}=c_a(\nabla^2+\frac14g^2\eta^2)\bar{P}^a+c(\nabla^2+\frac14g'^2\eta^2)\bar{P}+c_\al\nabla^2\bar{P}^\al+\text{vertices},
\end{gathered}
\end{equation}
where $\nabla_\mu$ is a covariant derivative and ``vertices'' denotes the terms of a higher power in the ghost fields. These terms do not contribute to the one-loop effective action.

\begin{table}
  \centering
  \begin{tabular}{|c|c|c|}
    \hline
    Name & Mass squared & \# \\\hline
    Higgs & $3\la\eta^2+\mu^2$ & $1\times1$ \\
    Goldstone Z & $\bar{m}^2+m_Z^2$ & $1\times1$ \\
    Goldstone W & $\bar{m}^2+m_W^2$ & $2\times1$ \\
    Z & $(g^2+g'^2)\eta^2/4$ & $1\times4$ \\
    W & $g^2\eta^2/4$ & $2\times4$ \\
    Ghosts to $SU(2)$ & $m^2_W$ & $-6\times1$ \\
    Photons & 0 & $1\times4$ \\
    Ghosts to $U(1)$ & $m^2_Z-m^2_W$ & $-2\times1$ \\
    Gluons & 0 & $8\times4$\\
    Ghosts to $SU(3)$ & 0 & $-16\times1$ \\
    \hline
  \end{tabular}\quad
  \begin{tabular}{|c|c|c|}
    \hline
    Name & Mass & \# \\\hline
    $u_i$ & $\dfrac{f_i^u}{\sqrt{2}}\eta$ & $9\times4$ \\
    $d_i$ & $\dfrac{f_i^d}{\sqrt{2}}\eta$ & $9\times4$ \\
    $e, \mu, \tau$ & $\dfrac{f_i^l}{\sqrt{2}}\eta$ & $3\times4$ \\
    $\nu_i$ & 0 & $3\times2$\\
    \hline
  \end{tabular}\quad
  \begin{tabular}{|c|c|}
    \hline
    Quantity & GeV \\\hline
    $\eta_0$ & 247 \\
    $m_{H0}$ & 129 \\
    $m_{Z0}$ & 91.2 \\
    $m_{W0}$ & 80.4 \\
    $m_{t0}$ & 171.2 \\
    $m_{b0}$ & 4.2 \\
    $m_{c0}$ & 1.27 \\
    $m_{\tau0}$ & 1.777 \\
    \hline
  \end{tabular}
  \caption{{\footnotesize The spectrum of particles of the standard model in the Feynman gauge. The bosons and ghosts to them are presented in the left table. The characteristics of the fermionic particles are given in the center table. For brevity, we introduced the notation $\bar{m}^2:=\la\eta^2+\mu^2$. The last column in these tables contains the number of degrees of freedom for each particle: the number of particles $N$ of a given type multiplied by the number of their polarizations in the Feynman gauge. The ghost fields have a negative number of degrees of freedom by definition. The right table collects the vacuum expectation value of the Higgs field and the masses of particles that are greater than $1$ GeV. The experimental data were taken from \cite{PDG} and correspond to the renormalization scale $2$ GeV. The recommended value of the Higgs boson mass is also given.}}\label{spectrum}
\end{table}

Now it is not difficult to deduce the spectrum of real and fictitious particles of the standard model in the Feynman gauge. It is presented in Table \ref{spectrum}. This is the only information about the standard model that we need for the one-loop calculations. Notice that all the masses of massive particles are proportional to $\eta$ except for the Higgs and Goldstone bosons. Furthermore, small deviations of $\eta$ result in the linear response of all the masses.

The Killing vector $\xi^\mu$ characterizing stationarity of the system can be straighten ($\xi^\mu=(1,0,0,0)$) by an appropriate coordinate change\footnote{We use the standard definition for stationary and static spacetimes (see, e.g., \cite{Wald}). The spacetime is called stationary if it possesses a Killing vector $\xi^\mu$ such that $\xi^2=g_{\mu\nu}\xi^\mu\xi^\nu>0$. The stationary spacetime is said to be (locally) static if $\nabla_{[\mu}(\xi_{\nu]}/\xi^2)=0$.}. Then the formal expression for the one-loop contribution of one bosonic mode to the effective action on a stationary background can be cast into the form
\begin{equation}\label{one-loop effact}
    \Ga^{(1)}_{1b}=-T\sum_k\frac{E^-_k}2,
\end{equation}
where $T$ is the time interval and $E^-_k$ is the energy of the mode $k$ corresponding to an antiparticle. This contribution is proportional to the zero-point energy and results from the normal ordering of the creation-annihilation operators. Another equivalent form for the one-loop contribution reads as
\begin{equation}\label{one-loop effact1}
    \Ga^{(1)}_{1b}=-T\Sp\int_0^\infty \frac{dp_0}2\theta(-G^{-1}(-p_0))=-T\im\Sp\int_0^\infty \frac{dp_0}{2\pi}\ln(G^{-1}(-p_0)),
\end{equation}
where $G(p_0)$ is the Fourier transform in $x^0$ of Green's function corresponding to the field considered on a curved background. The Heaviside step function $\theta(L-\la)$ of the Hermitian operator $L$ defines the spectral decomposition of unity associated with this operator. As for fermions, they give the same contributions but with an opposite sign in Eq. \eqref{one-loop effact}. The complete one-loop contribution to the effective action of the standard model on a stationary background is given by the sum of the contributions \eqref{one-loop effact} over all the species of particles multiplied by their number $N$ presented in Table \ref{spectrum}.

\section{Effective potential}\label{EffPot}

Now we are in position to calculate the one-loop effective action. To this end, we have to prescribe a rigorous meaning to the formal expressions \eqref{one-loop effact} or \eqref{one-loop effact1} and regularize them. One can distinguish two ways, at least, how to do this: we can employ the commonly used regularization methods based on the notion of a Fock proper-time \cite{Fock,Feyn,Schwing}, or introduce the energy cutoff -- the so-called physical regularization (see, e.g., \cite{Collins}). The first approach uses the heat kernel representation for the complex power of the Feynman propagator
\begin{equation}\label{propertime cutoff}
  G_F^\la=-\int_{-\infty}^0\frac{ids}{\Gamma(\la)}(is)^{\la-1}e^{-isG_F^{-1}},\qquad\la\in\mathbb{C}.
\end{equation}
As usual, the Feynman propagator $G_F$ is specified by the mass shift $m^2\rightarrow m^2-i\epsilon$. The one-loop correction can be expressed in terms of the trace of the operator $G^\la$ (see, e.g., \cite{DeWittDTGF,VasilHeatKer,AvramidiPhD}). The regularization is achieved by the analytic continuation in $\la$ (the zeta-function regularization), or in the parameter $d$ -- the dimension of the spacetime -- entering the heat kernel (the dimensional regularization), or by a mere cutoff on the upper integration limit. All these regularizations, which we shall collectively call the proper-time regularizations, give the same answer up to redefinition of the arising infinite constants. This is the general property of regularization schemes depending on one parameter. They are all equivalent in the aforementioned sense provided the multi-dimensional divergent integral (the trace, in our case) is reduced to the one-dimensional divergent integral over the same variable (the proper-time $s$, in our case). The regularization of one-dimensional integrals is uniquely defined up to the freedom mentioned above \cite{GSh}. As regards the multi-loop contributions, in the Feynman parameterization, they are all reduced to the one-dimensional integral with respect to the sum of the proper-times.

The second regularization prescription can be conveniently realized by introducing the Fermi-Dirac thermal cutoff. Namely, consider the one-loop $\Omega$-potential for fermionic fields at zero chemical potential (in the system of coordinates where $\xi^\mu=(1,0,0,0)$)
\begin{equation}\label{energy cutoff omeg}
  \Omega=\beta_0^{-1}\ln Z(\beta_0)=\Sp\int_0^\infty dp_0\frac{\theta(G^{-1}(-p_0))}{e^{\beta_0 p_0}+1}=\im\Sp\int_0^\infty \frac{dp_0}{\pi}\frac{\ln(-G^{-1}(-p_0))}{e^{\beta_0 p_0}+1},
\end{equation}
with $\beta_0$ being a reciprocal temperature defining the cutoff. Then the average energy takes a familiar form and reproduces the zero-point energy in the high-temperature limit
\begin{equation}\label{energy cutoff}
    E=-\partial_{\beta_0}(\beta_0\Omega)=\sum_k\frac{E^-_k}{e^{\beta_0 E^-_k}+1}\underset{\beta_0\rightarrow0}{\longrightarrow}\sum_k\frac{E^-_k}2.
\end{equation}
This representation allows us to employ the known high-temperature expansions for the partition function of fermions to obtain the one-loop effective action. Other regularization schemes introducing the energy cutoff into the model will lead to the same answer for the effective action up to redefinition of the infinite constants.

In this paper, we shall investigate the second method of regularization -- the energy cutoff -- for the following reasons (see also \cite{Collins}). The energy cutoff\footnote{Strictly speaking, we need to cut out a finite dimensional subspace from the one-particle Hilbert space. Only in this case do we remove all the divergencies. To this aim, we can confine the particle into a sufficiently large ``box'' making its spectrum discrete and then single out the subspace in the one-particle Hilbert space. This subspace spans the mode functions with the energies not higher than the energy cutoff.} has a clear physical interpretation in terms of particles: the dispersion laws of particles just deviate from the relativistic dispersion law at very high energies. This method is also natural from the quantum mechanical viewpoint as it is formulated in the quantum mechanical terms. It does not lead to states with a negative norm. This regularization procedure is minimal, i.e., we need not to introduce new entities into the model (as, for example, the proper-time) when the notions of energy and dispersion law are defined. However, this method possesses the well-known drawbacks: it violates the Lorentz- and gauge symmetries and requires a fine-tuning of the coefficients at the counterterms to restore these symmetries in the effective action (see, e.g., \cite{BogolShir,Sukhan}); it is awkward in comparison, for example, with the standard proper-time dimensional regularization. As opposed to the energy cutoff, the regularization procedures using the proper-time representation for the Feynman propagator preserve almost all the symmetries of the model since the proper-time is a Lorentz- and gauge invariant quantity. However, the concept of proper-time is purely classical and it cannot be defined in quantum mechanical terms without enlarging the number of degrees of freedom (introducing, for instance, one extra dimension for it).

On a flat background, these two approaches to regularization are equivalent (see, e.g., \cite{Collins,SvaSva,CoVaZe,Fulling,olopqfeces}) in the sense that we can adjust the coefficients at counterterms so as to make the effective actions equal. However, as we shall see, these two schemes \eqref{propertime cutoff} and \eqref{energy cutoff} give inequivalent results on a curved background. The divergent and finite parts of the effective actions will differ. At that, as for the finite part, this inequivalence is nonpolymomial in fields and cannot be removed by counterterms or field redefinitions.

The high-temperature expansions of one-loop contributions to the $\Omega$-potential on an arbitrary curved background are rather huge (see, e.g., \cite{DowKen}). That is why we restrict ourself to finding only the effective potential of the Higgs field. We neglect all the derivatives of the Higgs field. Furthermore, we also discard all the terms proportional to the derivatives of the metric field. The former approximation is adequate when the Higgs field changes slowly on the scales characterizing the metric field, while the latter approximation is appropriate when the background metric is approximately Ricci-flat. In this case, most of the relevant terms of the high-temperature expansion proportional to the derivatives of the metric vanish. The only possible nonvanishing combinations containing derivatives and having appropriate dimensions are (the total derivatives are also omitted, cf. \cite{VasilHeatKer}, Eqs. (4.27), (4.28))
\begin{equation}\label{derivative terms}
    \nabla_\mu\ln\sqrt{\xi^2}\nabla^\mu\ln\sqrt{\xi^2},\qquad R_{\mu\nu\rho\s}R^{\mu\nu\rho\s},\quad (\nabla_\mu\ln\sqrt{\xi^2}\nabla^\mu\ln\sqrt{\xi^2})^2,\quad\nabla_{[\mu}(\xi_{\nu]}/\xi^2)\nabla^{[\mu}(\xi^{\nu]}/\xi^2),
\end{equation}
where $\xi^2=g_{\mu\nu}\xi^\mu\xi^\nu$. The last term in \eqref{derivative terms} vanishes for static metrics. We discard these terms and shall obtain thereby the leading contribution to the derivative expansion of the effective action. As we shall see, this approximation is quite reasonable for the metrics of stars and macroscopic black holes.
Every derivative of the metric or Higgs field diminishes, effectively, the contribution of the term to the effective action by the factor $(m r)^{-1}$, where the massive parameter $m$ is of the order of $1-100$ GeV (electroweak scale) and $r$ is a distance from the gravitating body.

Bearing these assumptions in mind, it is easy to calculate the $\Omega$-potential \eqref{energy cutoff omeg}. The simplest way to do this is to use the representation for the trace of an operator in terms of its symbol (see, e.g., \cite{BerezMSQ,Und,BorNeuWal,Salcedo}). If we neglect all the derivatives of the fields, then the symbol of the operator entering \eqref{energy cutoff omeg} is given by
\begin{equation}
    \theta(g^{\mu\nu}(x)p_\mu p_\nu-m^2),
\end{equation}
for a scalar field. Hence, the partition function takes the form
\begin{equation}
\begin{split}
    \ln Z(\beta_0)&\approx\beta_0\int_0^\infty dp_0\int\frac{d\spx d\spp}{(2\pi)^3}\frac{\theta(g^{00}p_0^2-2g^{0i}p_0p_i+g^{ij}p_ip_j-m^2)}{e^{\beta_0p_0}+1}\\
    &=\beta_0\int_0^\infty dp_0\int\frac{d\spx d\spp}{(2\pi)^3}\frac{\theta\left((g^{00}-\bar{g}_{ij}g^{0i}g^{0j})p_0^2+g^{ij}p_ip_j-m^2\right)}{e^{\beta_0p_0}+1}\\
    &=\beta_0\int_0^\infty dp_0\int\frac{d\spx d\spp}{(2\pi)^3}\frac{\theta\left(p_0^2+g_{00}(g^{ij}p_ip_j-m^2)\right)}{e^{\beta_0p_0}+1}\\
    &=\int\frac{d\spx d\spp}{(2\pi)^3}\left(\frac{|g|}{g_{00}}\right)^{1/2}\ln\left(1+e^{-\be_0g_{00}^{1/2}\sqrt{\spp^2+m^2}}\right),
\end{split}
\end{equation}
where we have used the relations \cite{MTWeng,LandLifshCTF}
\begin{equation}
    \bar{g}_{ik}g^{kj}=\de_i^j,\qquad g_{00}\det\bar{g}_{ij}=g,\qquad g^{00}-\bar{g}_{ij}g^{0i}g^{0j}=(g_{00})^{-1}.
\end{equation}
Thus, the contribution of one bosonic mode to the effective potential has the form
\begin{equation}\label{one-loop effpot}
  -\Ga^{(1)}_{1b}=-T\int d\spx\sqrt{|g|}\partial_{\beta}\biggl[\int\frac{d\spp}{(2\pi)^3}\ln\left(1+e^{-\be\sqrt{\spp^2+m^2}}\right)\biggr]_{\beta_0\rightarrow0},
\end{equation}
where $\be:=\be_0\sqrt{\xi^2}$ is the Tolman reciprocal temperature. The contribution of the spin degrees of freedom just multiplies \eqref{one-loop effpot} by the corresponding factor. We have derived the well-known result \cite{Tolman,DowKen} that, in the leading quasiclassical order, the influence of gravity on thermal contributions reduces to the substitution of the temperature by the Tolman temperature. Now we can use the high-temperature expansion of the Fermi-Dirac $\Omega$-potential on a flat background (see, e.g., \cite{Kapusta,DolJack})
\begin{equation}\label{eff potential}
    -\Ga^{(1)}_{1b}=T\int d\spx\sqrt{|g|}\biggl[\frac{7\pi^2}{240\be_0^4\xi^4}-\frac{m^2}{48\be_0^2\xi^2}+\frac{m^4}{64\pi^2}\left(\ln\frac{\be_0^2 m^2\xi^2}{\pi^2}+2\ga+\frac12\right)\biggr]_{\be_0\rightarrow 0},
\end{equation}
where $\ga$ is the Euler constant. The same answer would be obtained if we used the point-splitting regularization procedure (\cite{DeWpaper}, Eq. (247)). The one-loop correction \eqref{eff potential} to the effective potential is generally covariant and represents the leading term in the derivative expansion of the effective action. The Killing vector entering the effective potential is not an external structure on the spacetime. It is determined by the metric through the Killing equations. Formula \eqref{eff potential} on a spherically symmetric background can be also derived using the explicit quasiclassical representation of the mode functions \cite{olephfsb}. The total one-loop contribution to the effective potential is obtained by summing over all particles.

If we employed the proper-time regularization we would obtain the same result \eqref{eff potential}, but with $\xi^2$ replaced by one. This fact is quite expectable as the rate of the proper-time of a particle moving along the Killing vector $\xi^\mu$ is proportional to $\sqrt{\xi^2}$:
\begin{equation}
    d s=\sqrt{\xi^2}dt,
\end{equation}
where $t$ is the variable dual to the energy (time). Loosely speaking, when we use the energy cutoff, the regularization parameter becomes blue-shifted (in comparison with the proper-time regularization) by the gravitational field. This observation suggests a simple way how to modify the results obtained by means of the proper-time regularization to the energy cutoff regularization. An accurate inspection of this observation will be given elsewhere.

So, as we see, there are three type of divergencies in the effective potential. They have to be canceled by adding appropriate counterterms to the initial action of the standard model. This, in essence, reduces to replacement of the infinite constants $\be_0^{-4}$, $\be_0^{-2}$, and $\ln\be_0$ by some finite quantities; and all other quantities entering the effective potential are replaced by their renormalized values. After that, we should impose certain normalization conditions in order to fix five unknown constants $\la$, $\mu^2$, $\tilde{\La}$, $\mathcal{A}_0$, and $\mathcal{A}_2$, where $\tilde{\La}$ is related to the cosmological constant, while $\mathcal{A}_0$ and $\mathcal{A}_2$ are the dimensional constants at $\xi^{-4}$ and $\xi^{-2}$, respectively.

The first normalization condition we choose is that the effective potential reduces to the well-known Coleman-Weinberg potential \cite{ColWein} for the flat spacetime, i.e., at $\xi^2=1$, and the masses of particles take their experimental values there. It is convenient to take the values of masses on the scale $2$ GeV (see Table \ref{spectrum}) where the standard model is in perturbative regime and the one-loop approximation is reasonable. This normalization condition implies
\begin{multline}\label{eff potential1}
    V=\tilde{\La}+\frac{\mu^2}2\eta^2+\frac{\la}4\eta^4-\frac1{64\pi^2}\biggl(\sum_fm_f^4\ln\frac{m_f^2\xi^2}{m_{f0}^2}-\sum_bm_b^4\ln\frac{m_b^2\xi^2}{m_{b0}^2}\biggr)\\
    +\mathcal{A}_0[(\xi^2)^{-2}-1]\biggl(\sum_f-\sum_b\biggr)+\mathcal{A}_2[(\xi^2)^{-1}-1]\biggl(\sum_fm_f^2-\sum_bm_b^2\biggr).
\end{multline}
where the indices $f$ and $b$ indicate statistics and $m_0$'s denote the values of the masses of the particles at $\xi^2=1$. The coupling constants $\mu^2$ and $\la$ are determined by the equations
\begin{equation}
    \mu^2=-\la\eta_0^2+\frac1{32\pi^2}\biggl(\sum_f\al_fm^2_{f0}-\sum_b\al_bm^2_{b0}\biggr),\qquad m^2_{H0}=3\la\eta_0^2+\mu^2,
\end{equation}
where $\al$'s are defined by $\de m^2=:\al\de\eta^2$. This system of equations can easily be solved exactly, but its solution is rather awkward. For a rough estimate, the term in parenthesis in the right-hand side of the first equation can be omitted.

As for the second normalization condition, we demand that the energy-momentum tensor following from \eqref{eff potential1} vanishes at $\xi^2=1$ (henceforward, we neglect the cosmological constant). It follows from this condition that
\begin{equation}\label{normal cond}
    \left.V(\xi^2,\eta^2)\right|_{\xi^2=1}=0,\qquad\left.\frac{\partial V(\xi^2,\eta^2)}{\partial\xi^2}\right|_{\xi^2=1}=0,
\end{equation}
where $\eta$ is taken at the minimum of the effective potential $V$. These equalities fix the constant $\tilde{\La}$ and relate the constants $\mathcal{A}_0$ and $\mathcal{A}_2$:
\begin{equation}
    \tilde{\La}=-\frac{\mu^2}2\eta^2_0-\frac{\la}4\eta^4_0,\qquad\mathcal{A}_0\biggl(\sum_f-\sum_b\biggr)=-\frac1{128\pi^2}\biggl(\sum_fm_{f0}^4-\sum_bm_{b0}^4\biggr)-\frac{\mathcal{A}_2}2\biggl(\sum_fm_{f0}^2-\sum_bm_{b0}^2\biggr).
\end{equation}
Thus, we ascertain in our particular case that the energy cutoff and the proper-time regularizations give equivalent results for the effective action on a flat background ($\xi^2=1$, in our case) and are inequivalent in a curved spacetime.

\begin{figure}[t]
\centering

\includegraphics*[width=7cm]{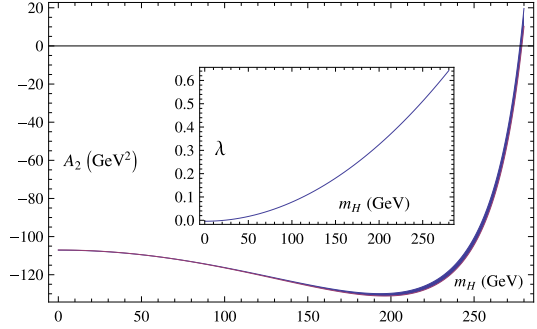}

\caption{{\footnotesize The dependence of the constant $\mathcal{A}_2$ on the Higgs boson mass according to the normalization condition \eqref{redshift_norm_cond}. Thickness of the line depicts the error related to the unknown constant $a$. Notice that $\mathcal{A}_2$ is almost uniquely determined and can be approximated by its value at $a=0$. The inset represents the dependence of the Higgs self-interaction coupling constant on the Higgs boson mass.}}
\label{A2}
\end{figure}

The only undetermined constant can be fixed imposing the third normalization condition. Namely, the deviation of the vacuum expectation value $\eta$ due to change of $\xi^2$ leads to the deviation of masses of all the massive particles and, as a consequence, to the deviation from the standard redshift law. Recall that in general relativity the redshift law is usually derived under the assumption that the emission spectrum does not depend on gravity in the reference frame associated with the emitter. However, if the masses of massive particles (for example, the electron mass) change with the gravitational field, the spectrum will also change even in this system of coordinates. It should be noted that the null redshift experiment \cite{Will} cannot apparently detect the mass-shift effect, at least in the linear order in gravitational perturbations. The additional corrections due to mass-shift to the emission spectra are the same as the changes of the linear sizes of the resonator in superconducting-cavity stabilized oscillator clocks. The sizes of the resonator vary in the same way as the Bohr radius of the hydrogen atom. So, the direct experiments are necessary. The direct tests of general relativity impose stringent constraints on the deviation from the standard redshift law in the weak field limit \cite{Will,Vessot}:
\begin{equation}
    \frac{\de\omega}{\omega}=(1+a)\vf_N,\qquad |a|<2\times10^{-4},
\end{equation}
where $\vf_N$ is the Newtonian potential and $\omega$ is the frequency of radiation. It is not difficult to brought this condition into the form
\begin{equation}\label{redshift_norm_cond}
    \left.\frac{\de\eta^2}{\eta_0^2}\right|_{\xi^2=1}=a\de\xi^2,
\end{equation}
where $\eta$ is taken at the minimum of the effective potential as before and $\de\xi^2\approx2\vf_N$. This condition fixes the last undetermined constant in the effective potential \eqref{eff potential1}. Making a negligible error, we can set $a=0$ (see Fig. \ref{A2}). Then
\begin{equation}\label{normal cond 2}
    \left.\frac{\partial V(\xi^2,\eta^2)}{\partial\xi^2\partial\eta^2}\right|_{\xi^2=1}=0\quad\Rightarrow\quad\mathcal{A}_2=-\frac1{32\pi^2}\frac{\sum_f\al_fm^2_{f0}-\sum_b\al_bm^2_{b0}}{\sum_f\al_f-\sum_b\al_b}.
\end{equation}
As a result, we can completely describe the gravitational mass-shift effect in the standard model on a stationary background.

\begin{figure}[t]
\centering

\includegraphics*[width=7cm]{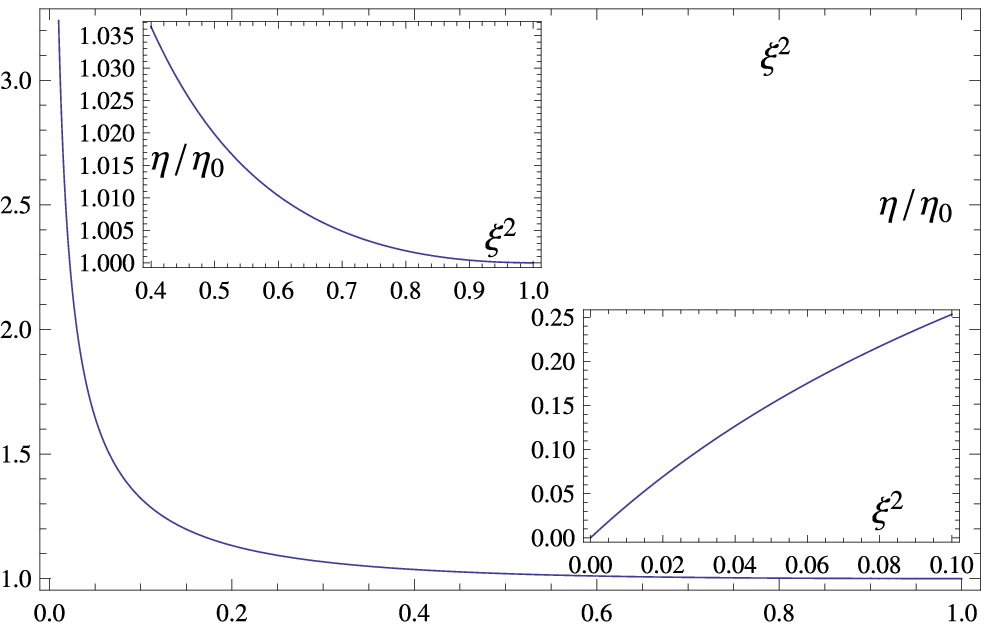}\;
\includegraphics*[width=7cm]{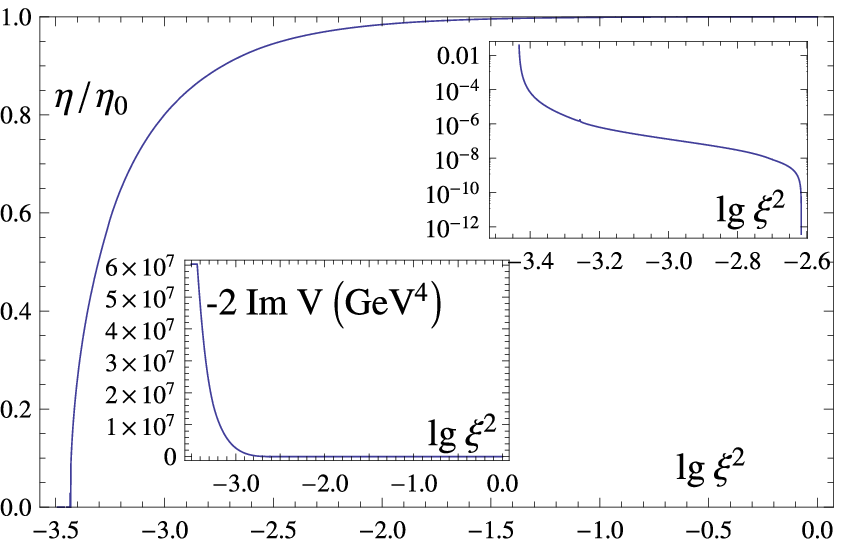}\\
\includegraphics*[width=7cm]{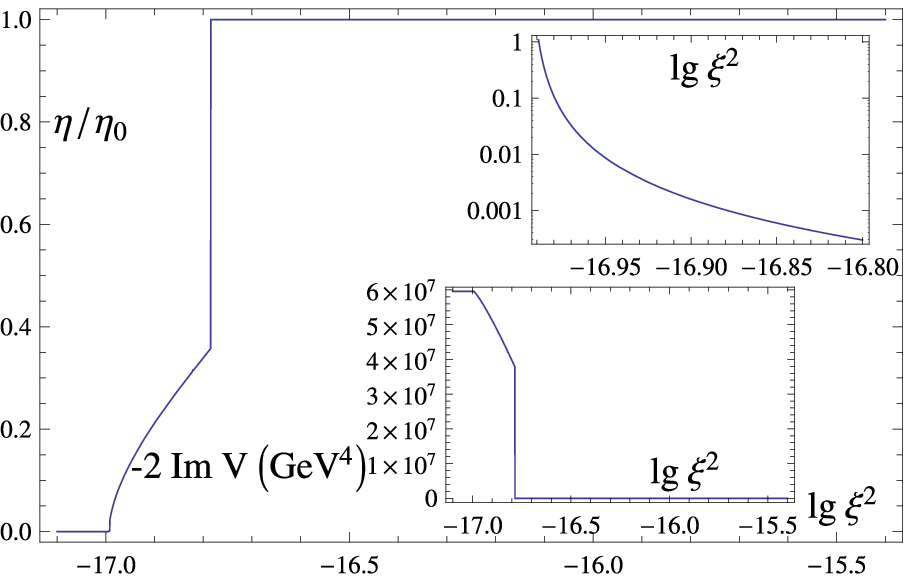}\;
\includegraphics*[width=7cm]{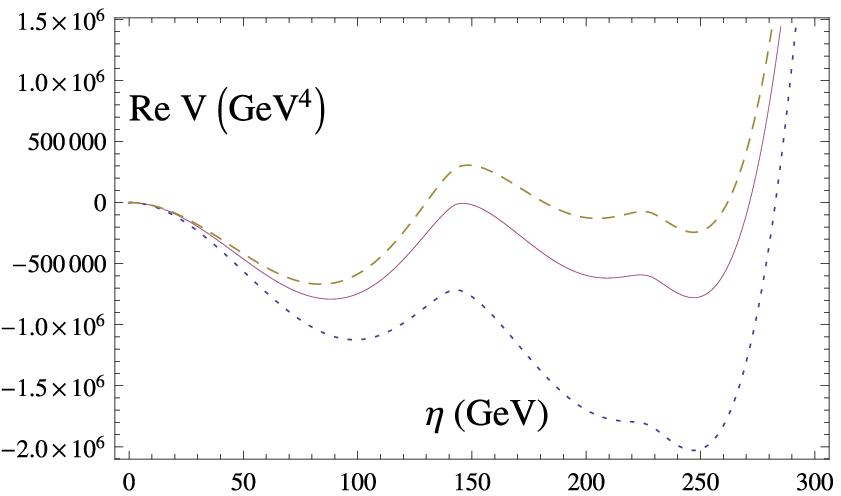}

\caption{{\footnotesize Left top panel: The dependence of the vacuum expectation value of the Higgs field on $\xi^2$ at the recommended value of the Higgs boson mass. The left inset represents the same dependence, but at smaller $\xi^2$. The right inset depicts the relative error $1-\eta_{as}/\eta$ of the asymptotics \eqref{eta_asympt}. Right top panel: The dependence of the vacuum expectation value of the Higgs field on $\lg\xi^2$ at $m_{H0}=279$ GeV. The left inset represents the dependence of the imaginary part of the Lagrangian doubled. The right inset provides the dependence of the quantity \eqref{appl_cond} on $\lg\xi^2$. This quantity characterizes the applicability of the in-out formalism in the case at issue. Left bottom panel: The same as for the right top panel, but for the critical value of the Higgs boson mass $m^{cr}_{H0}\approx278.2$ GeV. Notice that the masses do not almost change in this case provided $\xi^2$ is not extremely small. Right bottom panel: The shape of the real part of the effective potential is shown in a vicinity of the phase transition for the critical value of the Higgs boson mass (the value of the real part of the effective potential at $\eta=0$ is subtracted). The solid line corresponds to $\lg\xi^2=-16.785$, the dotted line is for $\lg\xi^2=-16.75$, and the dashed line represents the case $\lg\xi^2=-16.8$. It is clearly seen from these plots that the first phase transition is of the first order.}}
\label{MH}
\end{figure}

The standard model is not an accomplished theory. The Higgs boson has not yet been found and its mass is a free parameter of the model. Therefore several possible scenarios are presented in Fig. \ref{MH}. They are the same as for $O(N)$-$\phi^4$-model considered in \cite{olephfsb}, but ``inverted'', because the fermionic degrees of freedom dominate in the standard model. There is a critical value of the Higgs boson mass $m^{cr}_{H0}\approx278.2$ GeV where the constant $\mathcal{A}_2$ vanishes. When the Higgs boson mass is equal to or greater than $m^{cr}_{H0}$, the masses of all the massive particles decrease with the gravitational field and, eventually, go to zero at sufficiently small $\xi^2$. The broken symmetry of the standard model is restored. The system passes through the phase transition of the second order for $m_{H0}>m^{cr}_{H0}$. In case $m_{H0}=m^{cr}_{H0}$, it passes through the phase transition of the first order and then of the second order (see Fig. \ref{MH}). When the expectation value $\eta$ decreases, the masses squared of the Goldstone and Higgs bosons become negative. This leads to an appearance of the imaginary part in the effective potential and says that the vacuum state gets unstable. The values of the logarithms entering the effective potential taken on their cuts are uniquely specified by the prescription $m^2\rightarrow m^2-i\epsilon$. Strictly speaking, in this case we have to use the in-in formalism for systems with a non-stationary vacuum state. However, if the imaginary correction to the expectation value is rather small,
\begin{equation}\label{appl_cond}
    \left|\frac{\im V'_{\eta^2}(\xi^2,\eta^2)}{\re(\eta^2V''_{\eta^2\eta^2}(\xi^2,\eta^2))}\right|\ll1,
\end{equation}
we can still use the in-out effective action for the approximate evaluation of averages. The rate of decay of the vacuum state \cite{Schwing}, i.e., the probability, per unit time per unit volume, that a pair is created, is approximately given by the imaginary part of the effective Lagrangian doubled. In our case, it is nonnegative, as it should be, and is written as
\begin{equation}
    -2\im V=\frac1{32\pi}\left[m_H^4\theta(-m_H^2)+2m_{GW}^4\theta(-m_{GW}^2)+m_{GZ}^4\theta(-m_{GZ}^2)\right],
\end{equation}
where $m_{GW}$ and $m_{GZ}$ are the masses of the corresponding Goldstone bosons. The imaginary part arises when
\begin{equation}
    \eta^2/\eta_0^2\lesssim(1+2m_{W0}^2/m_{H0}^2)^{-1}.
\end{equation}
The numerical analysis shows that condition \eqref{appl_cond} is fulfilled for any $\xi^2$ at which the phase transition occurs and, of course, for larger $\xi^2$ (see Fig. \ref{MH}).

In the opposite case, i.e., when $m_{H0}\leq m^{cr}_{H0}$, and, in particular, for the recommended value of the Higgs boson mass $m_{H0}=129$ GeV, the masses of massive particles grow, when $\xi^2$ decreases, and tend to infinity at $\xi^2\rightarrow0$. At small $\xi^2$, the asymptotics of the expectation value of the Higgs field is given by
\begin{equation}\label{eta_asympt}
    \eta^2/\eta^2_0\approx c/\xi^2,
\end{equation}
where the dimensionless constant $c$ is found from the equation
\begin{multline}
    \biggl(\sum_f\al_f-\sum_b\al_b\biggr)\frac{2\mathcal{A}_2}{\eta_0^2}+\biggl\{\la-\biggl(\sum_f\al^2_f-\sum_b\al^2_b\biggr)\frac{1+\ln c^2}{32\pi^2}\\
    -\frac1{16\pi^2}\biggl[\al_H^2\ln\frac{m^2_{H0}}{\al_H\eta_0^2}+2\al_{GW}^2\ln\frac{m^2_{GW0}}{\al_{GW}\eta_0^2}+\al_{GZ}^2\ln\frac{m^2_{GZ0}}{\al_{GZ}\eta_0^2}\biggr]\biggr\}c=0.
\end{multline}
It has the solution $c\approx9.75\times10^{-2}$ at the recommended value of the Higgs boson mass. In Fig. \ref{MH}, this asymptotics is presented in comparison with the exact expectation value.

Some comments on these scenarios are in order. At small $\xi^2$, the quantum correction becomes large and the gradients of the fields $\xi^2$ and $\eta$ increase too. Therefore we need to take into account the higher-loop corrections, the gradient terms, and back-reaction when $\xi^2$ tends to zero. Moreover, the Higgs field self-interaction coupling is large at large the Higgs boson mass $m_{H0}$: $\la\approx0.63$ at the critical value of the Higgs boson mass. Hence, the critical value given above is just an estimate, and the scenarios described for large $m_{H0}$ are rather qualitative than quantitative. The lattice simulations and higher-loop calculations for the standard model at finite temperature nevertheless show that such one-loop results are quite reasonable (see, e.g., \cite{Kapusta,GynVeps}). In the next section we shall see that very small values of $\xi^2$ seem not to realize even for black holes. Also notice that the effective potential \eqref{eff potential1} evidently depends on the spectrum of masses of particles of the standard model. It will change if some new heavy particles are discovered.

\section{Non-stationary case}\label{NonStatCase}

In the previous sections, we considered the standard model on a stationary background when the Killing vector $\xi^\mu$ is defined. In the long run, the square of this vector entered the effective potential of the Higgs field. A natural question arises how to generalize the above results to non-stationary backgrounds. It turns out that this generalization is essentially unique under certain reasonable assumptions.

To investigate the dynamics of quantum field on a non-stationary background we have to use the in-in formalism and the in-in effective action (see \cite{SchwingBrown,Keldysh,DeWittDTGF,CalzHu,GFSh}). This approach doubles the number of the quantum and background fields $(g_{\mu\nu},\tilde{\Phi})\rightarrow(g^\pm_{\mu\nu},\tilde{\Phi}^\pm)$, where $\tilde{\Phi}=(\Phi,c,\bar{P})$. Further, we must prescribe the exact meaning to the operators entering the Heisenberg equations and introduce an appropriate regularization. The self-consistent definition of operators in the Fock space needs the definition of the normal ordering \cite{Lowenst,BogolShir,BerezMSQ}. Different prescriptions for the normal ordering lead to inequivalent theories on a curved background (see, e.g., \cite{DeWpaper}), while the formal manipulations with divergent operators may give rise to incorrect results like the absence of anomalies (whereas they are known to exist). On a stationary background, we have a preferred set (or sets) of the creation-annihilation operators and, consequently, a preferred prescription for the normal ordering. These creation-annihilation operators are associated with the stationary mode functions that are eigenfunctions of the Lie derivative with respect to $\xi^\mu$. In the case of a degenerate one-particle energy spectrum, there is a freedom to combine the mode functions corresponding to the same energy, but this freedom does not affect the observables. It just results in a unitary transform in the finite dimensional subspace of the one-particle Hilbert space provided the particle is confined into a large ``box''. Generalizing this construction to non-stationary backgrounds, we introduce the background vector field $\xi^\mu$ as well and require that this vector coincides with the Killing vector in the case of a stationary background. As before, it is normalized on unity for the flat spacetime: $\xi^2=1$. This vector field allows us to define the normal ordering and the physical regularization.

The regularized in-in effective action becomes the functional of the form
\begin{equation}
    \Ga[g^+_{\mu\nu},\xi^+_\mu,\tilde{\Phi}^+;g^-_{\mu\nu},\xi^-_\mu,\tilde{\Phi}^-].
\end{equation}
It is generally covariant with respect to the first and second arguments, separately. This implies
\begin{equation}\label{Ward_id}
    \nabla^\nu T_{\mu\nu}\approx\mathcal{L}_\xi\Ga_\mu+\nabla^\nu\xi_\nu\Ga_\mu,\qquad \sqrt{|g|}T^{\mu\nu}:=-2\left.\frac{\de\Ga}{\de g^+_{\mu\nu}}\right|_{+=-},\quad\sqrt{|g|}\Ga_\mu:=-\left.\frac{\de\Ga}{\de\xi^{+\mu}}\right|_{+=-}.
\end{equation}
Here the approximate equality means that the equations of motion for the fields $\tilde{\Phi}$ are taken into account and we identify the ``plus'' and ``minus'' fields upon variation. We see that covariant divergenceless of the energy-momentum tensor (the average of the energy-momentum tensor operator) is violated by the terms depending on the vector field $\xi^\mu$. In the stationary case, these terms disappear since $\xi^\mu$ is the Killing vector (see Eq. \eqref{Ward_id}), and the energy-momentum tensor is covariantly nondivergent. In the non-stationary case, we just impose this condition on the energy-momentum tensor and fix thereby the vector field $\xi^\mu$. Recall that the divergenceless of the energy-momentum tensor follows from the Einstein equations and so the latter requirement is a mere self-consistency condition (a Ward identity). The equations of motion for the vector field $\xi^\mu$ can be cast into the hydrodynamic form
\begin{equation}\label{eqmot xi}
    \nabla_\mu(\xi^\mu w)=0,\qquad \xi^\mu\nabla_{[\mu}(\Ga_{\nu]}/w)=\mathcal{L}_\xi(\Ga_{\nu}/w)=0,
\end{equation}
where $w:=\xi^\rho\Ga_\rho$. These equations should be supplemented by the initial and boundary conditions discussed above. It follows from the first equation that there exists a conserved charge in the system. Matching Eqs. \eqref{eqmot xi} with the equations of motion of the relativistic fluid (see, e.g., \cite{LandLifhyd}), it is natural to identify this charge with the entropy of the system. The second equation implies that if the $1$-form $\omega^{-1}\Ga_\mu$ was initially exact, it remains exact along the integral curves of the vector field $\xi^\mu$. In this case, the spacetime is foliated by the hypersurfaces associated with this integrable $1$-form. This provides a preferred definition of the energy of the system on a non-stationary background as a flux of the vector field $T^{\mu\nu}\xi_\nu$ through one of these hypersurfaces. Of course, this energy is not generally conserved as long as the background is non-stationary.

In a certain sense, the vector field $\xi^\mu$ spoils the so-called local position invariance (see, e.g., \cite{Will}). Hence, the approach we are considering can be placed among the Lorentz-invariance violating theories (for a review, see \cite{Kostel,Odints}). Notice, however, a distinction of this approach with the standard Lorentz-invariance violating models like, for example, the minimal Lorentz-invariance violating standard model extension \cite{Kostel}. The vector field $\xi^\mu$ we have introduced is always a background field, i.e., it does not represent a quantum field or its average. It characterizes the regularization procedure and is necessary for a proper definition of the composite field operators in the Hilbert space. Whatever new particles are introduced into the theory, this vector field has to be included when it comes to define the physical regularization. Therefore its dynamics do not follow from the action principle, at least immediately, but arise as a self-consistency condition. Moreover, this approach is minimal: i) There is no Lorentz-invariance violation for a flat background; ii) There is no new additional structures or fields in the effective action on a stationary background. The vector field $\xi^\mu$ is just the Killing vector of the metric. In the non-stationary case, this vector field is uniquely determined by the self-consistency condition \eqref{eqmot xi} with the initial and boundary conditions described above; iii) The number of new parameters entering the effective action is minimal, which allows us to make predictions rather than just to fit the model to the experimental data.

As an example, we consider the Schwarzschild black hole. Despite the fact that it is described by a static metric except for the small region near the horizon where the matter accretes, it should be thought of as an infinitely lasting collapse, i.e., as a non-stationary system. Later, this observation will be relevant for the analysis of the solutions to \eqref{eqmot xi}.

Now we should make certain approximations in order to solve the equation of motion for the vector field $\xi^\mu$. First, we assume that the system is in a vacuum state, i.e., there are no particles in it, or their back-reaction on the vacuum is marginal, and the Hawking particle production \cite{Hawk} is also negligible for the averages of quantum fields. Then the in-in effective action entering \eqref{eqmot xi} can be approximated by the in-out effective action for the vacuum:
\begin{equation}
    \Ga[g^+_{\mu\nu},\xi^+_\mu,\tilde{\Phi}^+;g^-_{\mu\nu},\xi^-_\mu,\tilde{\Phi}^-]\approx\Ga[g^+_{\mu\nu},\xi^+_\mu,\tilde{\Phi}^+]-\Ga^*[g^-_{\mu\nu},\xi^-_\mu,\tilde{\Phi}^-].
\end{equation}
Second, we use the same assumptions which we made in deriving the effective potential in the previous section. That is, we replace the effective action entering \eqref{eqmot xi} by the effective action constructed from the effective potential \eqref{eff potential1}. Then we have
\begin{equation}\label{eqmot xi app}
    \Ga_\mu=2\xi_\mu\frac{\partial V(\xi^2,\eta^2)}{\partial\xi^2},\qquad w=2\xi^2\frac{\partial V(\xi^2,\eta^2)}{\partial\xi^2},\qquad \Ga_\mu/w=\xi_\mu/\xi^2,
\end{equation}
where $\eta$ is taken at the minimum of the effective potential. The last relation in \eqref{eqmot xi app} holds in a general case provided that the vector field $\xi^\mu$ appears in the in-in effective action as $\xi^2$. The combination $\xi_\mu/\xi^2$ is nothing but the Tolman reciprocal temperature $1$-form when $\xi^\mu$ is the Killing vector. It is closed when the metric is static and, hence, it is exact provided the fundamental group of the spacetime is trivial. So, if the system starts its evolution from the state with a static metric, the $1$-form $w^{-1}\Ga_\mu$ will be exact over all the spacetime in this case.

The equations of motion \eqref{eqmot xi} together with \eqref{eqmot xi app} are exactly the equations describing a hot ultrarelativistic fluid \cite{LandLifhyd},
\begin{equation}
    \be:=\sqrt{\xi^2},\qquad\s:=\beta w,\qquad p:=-V,
\end{equation}
being the reciprocal temperature, the entropy density, and the pressure, respectively, and $w=:\e+p$ being the enthalpy density (for the other hydrodynamic descriptions of a vacuum, see, e.g., \cite{ChefNov}). As usual, the energy density is denoted by $\e$. The temperature scale is chosen such that the temperature $(\xi^2)^{-1/2}$ is equal to unity for the flat spacetime. As a result, this temperature is dimensionless and the entropy density has the dimension of the energy density. These quantities can be also obtained from the standard definition of the energy-momentum tensor if we vary the effective action with the potential $V(\xi^2,\eta^2)$ with respect to the metric. The normalization conditions \eqref{normal cond} imply that the entropy, enthalpy, and energy densities, as well as the pressure of the vacuum state vanish for the flat spacetime.

In the stationary case, we have already seen that the physical regularization can be achieved by introducing a thermal cutoff as if the system were at the very high temperature characterized by the energy cutoff. Therefore it is not surprising that, in the non-stationary case, a vacuum acquires the properties of an ultrarelativistic fluid. According to this interpretation, one can think of the particle-antiparticle virtual pairs as some particles created at the expense of the energy of a thermostat heated up to the cutoff temperature. This energy is given back to the thermostat when these particles annihilate. All the charges are conserved during this process and the average energy is also conserved in the stationary case. Of course, this is just one of the possible ways of thinking of the vacuum and particle-antiparticle virtual pairs likewise the notion of Dirac's ``sea''. One may say that this sea is heated up to the cutoff temperature.

Inasmuch as the vacuum possesses the same properties as an ultrarelativistic fluid, we can define, in particular, the speed of sound in it
\begin{equation}\label{sound_sp}
    c_s^2:=\frac{\partial p}{\partial\e}=-\left(\frac{d\ln\s}{d\ln\be}\right)^{-1},
\end{equation}
i.e., the speed of propagation of small perturbations of the vector field $\xi^\mu$. Bearing in mind the normalization condition \eqref{normal cond}, we can easily find the expression for this speed in the weak field limit $\xi^2\rightarrow1$:
\begin{equation}\label{sound_sp_weak}
    c_s^2\approx\frac{1-\xi^2}{2}\approx-\vf_N,
\end{equation}
where we have assumed that
\begin{multline}
    \left.\sigma'(\xi^2)\right|_{\xi^2=1}=-2\left.p''(\xi^2)\right|_{\xi^2=1}\\
    =\frac{-1}{16\pi^2}\biggl[\sum_fm^4_{f0}-\sum_bm^4_{b0}-\biggl(\sum_fm^2_{f0}-\sum_bm^2_{b0}\biggr)\frac{\sum_f\al_fm^2_{f0}-\sum_b\al_bm^2_{b0}}{\sum_f\al_f-\sum_b\al_b}\biggr]
\end{multline}
is not zero. The last formula is easily derived if one takes into account the normalization condition \eqref{normal cond 2}. It is remarkable that the speed of sound is independent of details of the standard model in the weak field limit. It is solely determined by the Newtonian potential $\vf_N$ and equals the so-called first cosmic velocity. One may say that the virial theorem is fulfilled for this fluid. In the case when $\s'(1)=0$, while $\s''(1)\neq0$, the sound speed squared in the weak field limit is given by \eqref{sound_sp_weak} divided by $2$. This occurs at the Higgs boson mass $\bar{m}_{H0}\approx263.6$ GeV, but it is hard to imagine that this degenerate case is realized indeed. It is this case when the components of the energy-momentum tensor of the vacuum tend to zero as $1/r^2$ and not as $1/r$, where $r$ is a distance from the gravitating object.

At small $\xi^2$, the massless contribution to the entropy density dominates. This is not difficult to see from the asymptotics \eqref{eta_asympt} when the masses of particles grow with decreasing $\xi^2$. In the opposite case ($\mathcal{A}_2\geq0$), the masses tend to zero and, of course, the massless contribution also dominates. In both cases,
\begin{equation}\label{sound_sp_strong}
    \s\approx\beta^{-3}\biggl[\frac{1}{32\pi^2}\biggl(\sum_fm^4_{f0}-\sum_bm^4_{b0}\biggr)+2\mathcal{A}_2\biggl(\sum_fm^2_{f0}-\sum_bm^2_{b0}\biggr)\biggr],\qquad c_s\approx\frac1{\sqrt{3}},
\end{equation}
as one would expect for a gas of massless particles.

\begin{figure}[t]
\centering

\includegraphics*[width=7cm]{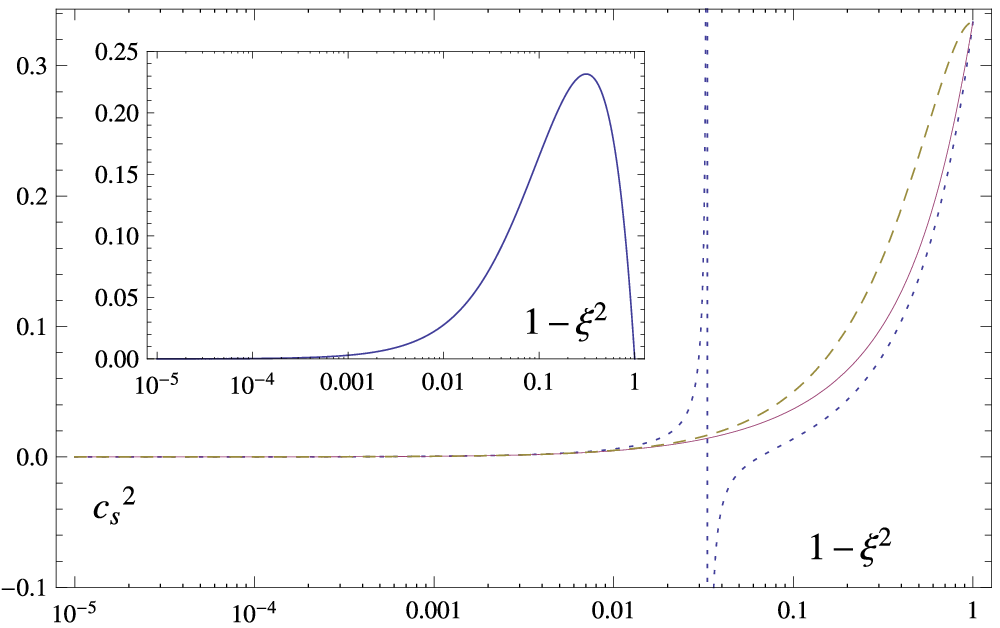}\;
\includegraphics*[width=7.7cm]{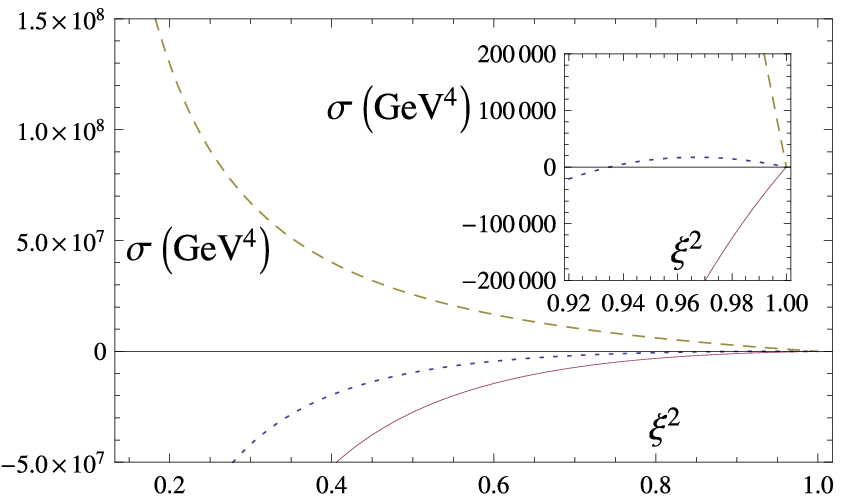}

\caption{{\footnotesize Left panel: The dependence of the sound speed squared on $1-\xi^2$. The solid line represents the sound speed at the recommended value of the Higgs boson mass. The dashed line corresponds to the critical value of the Higgs boson mass $m^{cr}_{H0}$. This plot coincides with the plot of the function \eqref{entr_sound_speed_A2}. The dotted line depicts the sound speed squared at the Higgs boson mass $m_{H=0}\approx264.6$ GeV belonging to the interval of hydrodynamical instability $(\bar{m}_{H0},m^{cr}_{H0})$. The inset represents the plot of the relation $c_{s0}/c_s-1$, where $c_{s0}$ is given by formula \eqref{entr_sound_speed_A2} and $c_s$ is the sound speed at the recommended value of the Higgs boson mass. Right panel: The dependence of the entropy density of the vacuum state on the reciprocal Tolman temperature squared $\xi^2$. The solid, dotted, and dashed lines correspond to the same cases as on the left panel. The inset depicts the same dependence, but for smaller $\xi^2$. We see that the entropy density is positive in the case $\mathcal{A}_2\geq0$ and is mostly negative otherwise. The entropy density changes its sign with $\xi^2$ at the Higgs boson masses lying in the ``instability'' interval $(\bar{m}_{H0},m^{cr}_{H0})$. In particular, the entropy density is nonpositive at the recommended value of the Higgs boson mass.}}
\label{SS}
\end{figure}

As seen from Fig. \ref{MH}, the masses of particles at the critical value of the Higgs boson mass ($\mathcal{A}_2=0$) do not almost change down to extremely small values of $\xi^2$. This allows us to find exact expressions for the entropy density and the speed of sound for sufficiently large $\xi^2$:
\begin{equation}\label{entr_sound_speed_A2}
    \s=\biggl(\sum_fm^4_{f0}-\sum_bm^4_{b0}\biggr)\frac{\beta^{-3}-\beta}{32\pi^2},\qquad c_s^2=\frac{1-\xi^4}{3+\xi^4}.
\end{equation}
Note that this speed does not depend on the parameters of the standard model. Its weak field limit coincides with \eqref{sound_sp_weak}, and it turns into \eqref{sound_sp_strong} at $\xi^2\rightarrow0$. The plots of $c_s^2$ versus $\xi^2$ are presented in Fig. \ref{SS} for different values of the Higgs boson mass. In the interval of the Higgs masses $m_{H0}\in(\bar{m}_{H0},m^{cr}_{H0})$, there is a range of the values of $\xi^2$ where the sound speed squared is negative and the entropy density changes its sign (see Fig. \ref{SS}). This signalizes that the system becomes hydrodynamically unstable, i.e., small fluctuations of the vector field $\xi^\mu$ grow exponentially with time. Since this occurs at large values of the Higgs boson mass $m_{H0}$, such a behavior may be a mere artefact of the approximations made and could be cured by higher-loop and derivative corrections to the effective action. However, this question needs a further investigation. Henceforth, we assume that the Higgs boson mass does not fall into this interval and the system considered is hydrodynamically stable.

Now we consider in detail the solution to the equations of motion \eqref{eqmot xi} for the Schwarzschild black hole. Upon the approximations made, we have just to describe a spherically symmetric accretion of an ultrarelativistic fluid onto a black hole. This is a well-studied subject (see, e.g., \cite{Bondi,Michel,ShapTeukol,BabDokEro}) and we consider only the main steps. In this case, the self-consistency condition \eqref{eqmot xi} is reduced to two equations
\begin{equation}\label{accretion eqs}
    1-\frac{r_g}{r}+u^2=\xi^2,\qquad r^2u\s(\xi^2)=k,
\end{equation}
where $r_g$ is the gravitational radius, $k$ is some constant characterizing the entropy flux, and $u:=\xi^r/\sqrt{\xi^2}$ is a radial component of the $4$-velocity. We have also used the normalization condition $\xi^2\rightarrow1$ at spatial infinity. The system of equations \eqref{accretion eqs} has the trivial solution
\begin{equation}
    k=0,\qquad u=0,\qquad \xi^2=1-\frac{r_g}{r}.
\end{equation}
It corresponds to the static system studied in the previous section or, physically, to a stable star (in a certain approximation, of course). Stability of the star means here that it does not appreciably change its gravitational field on the time scale of the order of $L/c_s$, where $L$ is a characteristic size of the star. A black hole represents an infinitely lasting collapse and so the entropy flux and the radial velocity $u$ are not zero in this case. We see that the Tolman reciprocal temperature $\beta$ for a black hole is greater than the reciprocal temperature for a static star at the same coordinate $r$. In particular, the reciprocal temperature is not equal to zero at $r=r_g$ for a black hole. Therefore the gravitational mass-shift effect is greater for a static star rather than for a black hole at the same $r$.

The system of equations \eqref{accretion eqs} has a unique nontrivial ($k\neq0$) nonsingular solution $u(r)$ satisfying the boundary condition
\begin{equation}\label{bound_cond}
    \lim_{r\rightarrow\infty}u=0.
\end{equation}
It possesses the weak field asymptotics
\begin{equation}
    u\approx-\frac{k}{\s'(1)r_gr},
\end{equation}
provided $m_{H0}\neq\bar{m}_{H0}$. The constant $k$ is uniquely defined by the regularity condition. A regular solution $u(r)$ should pass through the so-called critical point where, roughly speaking, the velocity of the accretion flow becomes equal to the speed of sound in it:
\begin{equation}\label{crit point}
    u^2_*=\xi^2_*c_s^2(\xi^2_*)=\frac{r_g}{4r_*}.
\end{equation}
The asterisk distinguishes the quantities taken at the critical point. The equations for the critical point follow from \eqref{accretion eqs}. If we differentiate \eqref{accretion eqs} then
\begin{equation}
    (\s+2u^2\s')\frac{du}{u}+(2r\s+r_g\s')\frac{dr}{r^2}=0.
\end{equation}
The quantity,
\begin{equation}\label{quantity}
    \s+2u^2\s'=\s\left(1-\frac{u^2}{\xi^2c_s^2}\right),
\end{equation}
changes its sign when $r$ runs from $r_g$ to $+\infty$ since
\begin{equation}
    \s\left(1-\frac{u^2}{\xi^2c_s^2}\right)\underset{r\rightarrow r_g}{\longrightarrow}\s(1-c_s^{-2}),\qquad\s\left(1-\frac{u^2}{\xi^2c_s^2}\right)\underset{r\rightarrow \infty}{\longrightarrow}\s.
\end{equation}
The entropy density has the same sign for any $\xi^2\in(0,1)$ and the sound speed is always less than unity provided the Higgs boson mass for the flat spacetime does not fall into the ``instability'' interval $(\bar{m}_{H0},m^{cr}_{H0})$. The requirement that the radial velocity has a finite derivative at the point, where the quantity \eqref{quantity} vanishes, leads to equations \eqref{crit point} for the critical point. Equations \eqref{crit point} allow us to find $r_*$, $u_*$, and $\xi^2_*$, and, hence, the constant $k$ from the second equation in \eqref{accretion eqs}.

\begin{figure}[t]
\centering

\includegraphics*[width=7cm]{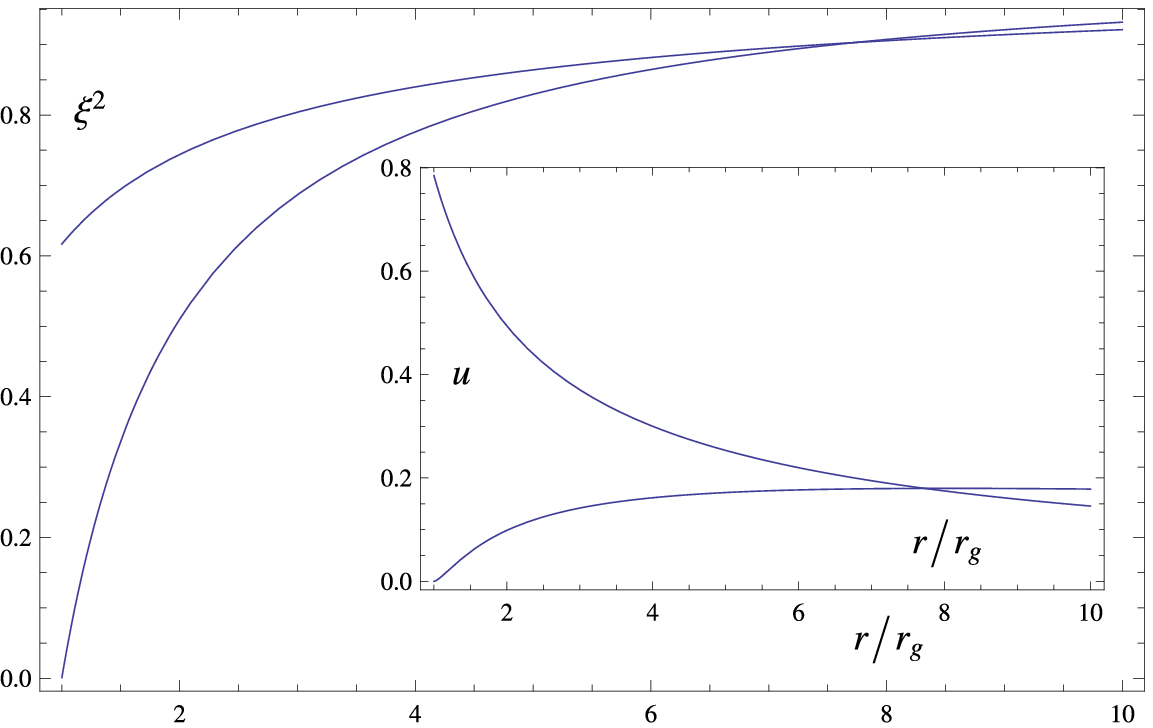}\;
\includegraphics*[width=7cm]{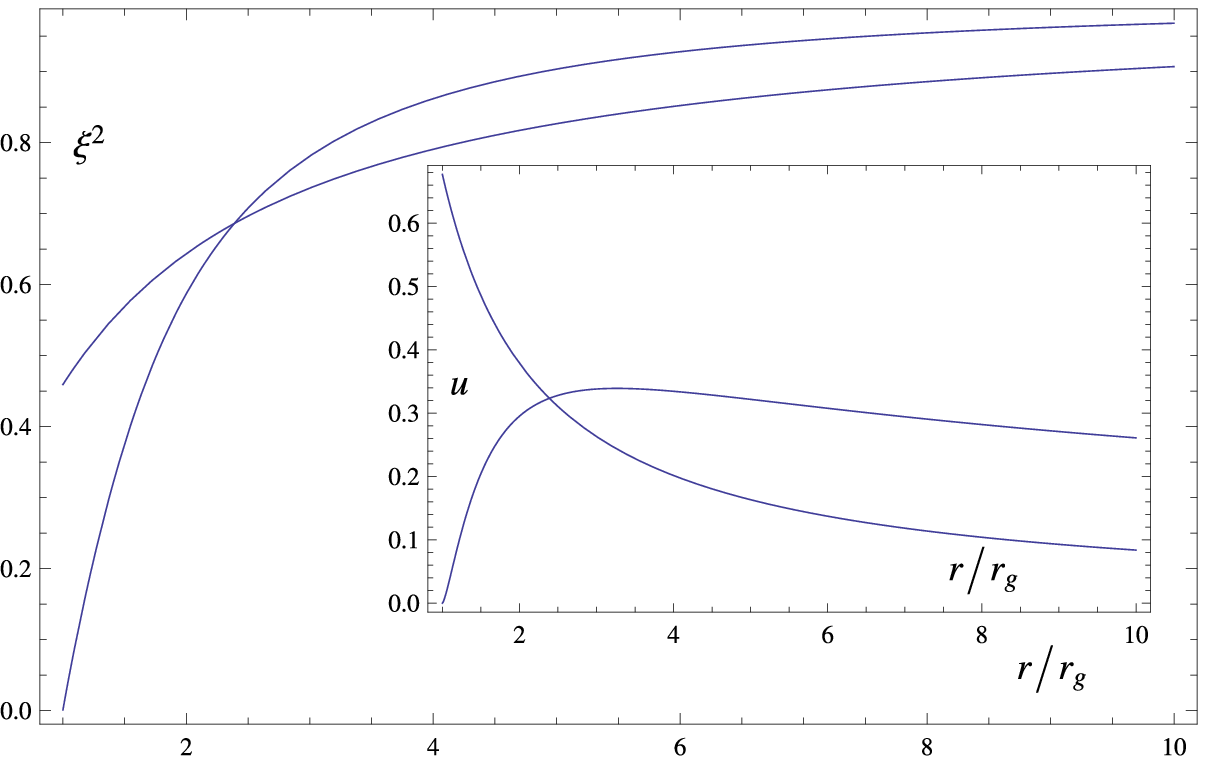}

\caption{{\footnotesize  The dependence of $\xi^2$ on the Schwarzschild coordinate $r$. Two branches of the solution to the accretion equations \eqref{accretion eqs} are presented. One branch corresponds to the ``infall'' solution, another one being the ``outflow'' solution. These branches intersect at the critical point. The insets depict the moduli of the accretion velocities. Only that branch of the general solution which does not vanish on the horizon represents the solution satisfying the boundary condition \eqref{bound_cond}. This is the infall solution. The case of the recommended Higgs boson mass is presented on the left panel and the case of the critical Higgs boson mass $m^{cr}_{H0}$ is given on the right panel. These plots together with the plots in Fig. \ref{MH} provide a complete description of the gravitational mass-shift effect for a Schwarzschild black hole.}}
\label{ACCR}
\end{figure}

Unfortunately, the critical point cannot be found analytically with the exception of the case where the Higgs boson mass $m_{H0}$ is equal (or close) to the critical mass $m^{cr}_{H0}$. However, the accretion problem can be easily analyzed numerically. The plots of the accretion velocity and the Tolman temperature are presented in Fig. \ref{ACCR} for the different Higgs boson masses. The critical point for the recommended Higgs boson mass is found to be
\begin{equation}
    \xi^2_*\approx0.90,\qquad u_*\approx-0.18,\qquad r_*/r_g\approx7.73,\qquad k/r_g^2\approx1.02\times10^7 \text{GeV}^4.
\end{equation}
As far as the critical value of the Higgs boson mass is concerned, we know explicit expressions for the entropy density and the sound speed \eqref{entr_sound_speed_A2} in this case. Substitution of the sound speed to the equations for the critical point gives the result
\begin{equation}
    \xi^2_*=\frac{\sqrt{33}-3}{4},\qquad u^2_*=\frac{7-\sqrt{33}}{12},\qquad r_*=\frac{3r_g}{7-\sqrt{33}}.
\end{equation}
Then the equation for the entropy flux conservation takes the form
\begin{equation}
    r_*^2u_*(\xi_*^{-3}-\xi_*)=\frac{r_g^2}8\sqrt{\frac32(63+11\sqrt{33})}=r^2\sqrt{\xi^2-1+r_g/r}(\xi^{-3}-\xi).
\end{equation}
This equation can be reduced to the fourth order polynomial equation on $1/r$. Only one root of this equation has physical meaning and satisfies the boundary condition \eqref{bound_cond}. The explicit expression for $r(\xi^2)$ is rather huge and we do not write it here, but it is remarkable that the accretion problem admits of an analytical solution.

We see from the plots presented in Fig. \ref{ACCR} that the singularities arising in the effective potential at $\xi^2=0$ are not actually realized. In particular, when $\mathcal{A}_2=0$, the minimal value of the reciprocal temperature squared, which it takes on the horizon $r=r_g$, becomes
\begin{equation}\label{xi2 on hor A2}
    \left.\xi^2\right|_{r=r_g}=\frac1{32}\left(\sqrt{1402+66\sqrt{33}}-\sqrt{378+66\sqrt{33}}\right)\approx0.46.
\end{equation}
As for the recommended Higgs boson mass, the numerical analysis shows that $\xi^2\approx0.62$ at $r=r_g$. This corresponds to the relative change of the vacuum expectation value $\de\eta/\eta_0\approx9.2\times10^{-3}$. Taken on the horizon, the Tolman reciprocal temperature squared $\xi^2$ decreases, when the Higgs boson mass $m_{H0}$ increases till $m_{H0}\approx232$ GeV. The reciprocal temperature squared is equal approximately to $0.56$ at this value of the Higgs boson mass. Then it increases up to unity at $\bar{m}_{H0}$. After the instability interval, the reciprocal temperature squared increases with $m_{H0}$ starting from \eqref{xi2 on hor A2}. At very large Higgs boson masses, applicability of the perturbation theory and the one-loop approximation becomes questionable and so we exclude this region from our consideration.

\section{Discussion}

In this paper we obtained the one-loop effective potential of the Higgs field both for stationary and non-stationary gravitational backgrounds. The crucial point was the use of the energy cutoff regularization or the normal ordering in prescribing a rigorous meaning to the Hamiltonian of the standard model and to the other composite operators. Having imposed  physically reasonable normalization conditions, we completely specified the effective potential and described thereby the vacuum state at the one-loop level. This, in turn, allowed us to obtain a complete description of the gravitational mass-shift effect on the Schwarzschild background in a certain approximation.

In particular, it appeared that the gravitational mass-shift effect is greater for static stars than for Schwarzschild black holes. For the non-rotating neutron star with the radius of $2r_g$ this effect gives the relative mass-shift $\de\eta/\eta_0\approx2.0\times10^{-2}$ (for the recommended value of the Higgs boson mass) on the surface of the star, while for the black hole it leads to $\de\eta/\eta_0\approx9.2\times10^{-3}$ on the horizon. A proper generalization of the approach to the non-stationary case allowed us to get rid of the divergences of the effective potential on the black hole horizon.

Besides, the properties of the vacuum proved to be similar to the properties of an ultrarelativistic fluid. It possesses the entropy and enthalpy densities, the pressure etc. The entropy density and the pressure of the vacuum turn out to be negative, when the Higgs boson mass is less than $263.6$ GeV, and they become positive for the Higgs boson masses greater than the critical value $278.2$ GeV. This implies the existence of a small screening of a gravitating object in the former case and a small anti-screening in the latter case. Although these effects are rather small, they become relevant on cosmic scales. The vacuum energy density tends to zero as $-\s'(1)r_g/r$ and the pressure behaves like $p''(1)r_g^2/2r^2$ at sufficiently large distance $r$ from the object. Therefore the energy of the vacuum state diverges in the limit of infinite space. Of course, this contribution to the vacuum energy is only relevant on the distances less than $|\s'(1)|r_g/\La$ as the contribution from the cosmological constant $\La$ dominates above this scale.

The results of this paper can be generalized in several directions. It would be interesting to investigate the properties of the vacuum state for the Kerr-Newman background. A naive substitution of the Kerr-Newman metric to the effective potential \eqref{eff potential1} gives rise to the divergence on the ergosphere. This problem seems to be resolved by solving the self-consistency condition \eqref{eqmot xi} as we have done for the Schwarzschild black hole. Another evident generalization is to include back-reaction and derivative corrections to the one-loop effective action, although, to all appearance, these corrections are small for macroscopic gravitating objects. It is also interesting to investigate the loop corrections to other observables, such as the electron form factors on a curved background using the energy cutoff. An evident generalization of the flat spacetime results suggested by the form of the effective potential \eqref{eff potential1} consists in replacement of the massive parameter $\mu$ of the dimensional regularization (or the cutoff parameter) by its blue-shifted counterpart $\mu(\xi^2)^{-1/2}$ provided the derivatives of $\xi^2$ are negligible. This will lead to small variations of the coupling constants with gravity (see for recent tests, e.g., \cite{Blatt}). However, this guess needs a further exploration. When this problem will be solved, the higher loop corrections to the effective potential of the Higgs field on a curved background can be studied making use of the energy cutoff regularization.

\begin{acknowledgments}

The work is supported by the RFBR grant 09-02-00723-a. I appreciate the anonymous referees for valuable comments. I am also grateful to Professor A.~A. Sharapov for careful reading of the manuscript.

\end{acknowledgments}

\end{document}